\documentclass{CFD2017-mod}
\usepackage{CFD2017-mod}
\usepackage{bm}
\usepackage{mathtools}
\usepackage{paralist}
\usepackage{siunitx,booktabs}
\usepackage{subfig}

\newcommand{\Fig}[1]{Figure \ref{fig:#1}}
\newcommand{\Tab}[1]{Table \ref{tab:#1}}
\newcommand{\Eq}[1]{Eq. (\ref{eq:#1})}
\newcommand{\bVec}[1]{\bm{#1}}

\title{IMPLEMENTATION, DEMONSTRATION AND VALIDATION OF A USER-DEFINED WALL FUNCTION FOR DIRECT PRECIPITATION FOULING IN ANSYS FLUENT}
\shorttitle{USER-DEFINED WALL FUNCTION FOR DIRECT PRECIPITATION FOULING IN ANSYS FLUENT}
\paperID{CFD 2017}
\author{Sverre}{G. Johnsen} 
\presenting  
\address{SINTEF Materials and Chemistry, NO-7465 Trondheim, NORWAY}
\email{sverre.g.johnsen@sintef.no}
\author{Tiina}{M. P\"{a}\"{a}kk\"{o}nen}
\address{University of Oulu, Environmental and Chemical Engineering, FI-90014 Oulu, FINLAND}
\author{Stein}{T. Johansen}
\sameaddressas{1}
\address{NTNU, Dept. of Energy and Process Engineering, NO-7491 Trondheim, NORWAY}
\author{Riitta}{L. Keiski}
\sameaddressas{2}
\author{Bernd}{Wittgens}
\sameaddressas{1} 

\begin{document}
\maketitle  
\headers   

\abstract{
In a previous paper \citep{Johnsen15} and presentation \citep{Johnsen16}, we developed and demonstrated a generic modelling framework for the modelling of direct precipitation fouling from multi-component fluid mixtures that become super-saturated at the wall. 
The modelling concept involves the 1-dimensional transport of the fluid species through the turbulent boundary layer close to the wall. 
The governing equations include the Reynolds-averaged (RANS) advection-diffusion equations for each fluid species, and the axial momentum and energy equations for the fluid mixture.
The driving force for the diffusive transport is the local gradient in the species' chemical potential. 
Adsorption mechanisms are not modelled per se, but the time-scale of adsorption is reflected in the choice of Dirichlet boundary conditions for the depositing species, at the fluid-solid interface.

In this paper, the modelling framework is implemented as a user-defined function (UDF) for the CFD software ANSYS Fluent, to act as a wall boundary condition for mass-transfer to the wall. 
The subgrid, 1-dimensional formulation of the model reduces the computational cost associated with resolving the fine length-scales at which the boundary-layer mass transfer is determined, and allows for efficient modelling of industry-scale heat exchangers suffering from fouling.

The current paper describes the modelling framework, and demonstrates and validates its applicability in a simplified 2D heat exchanger geometry (experimental and detailed CFD modelling data by \citet{Paakkonen12,Paakkonen16}). 
By tuning the diffusivity, only, good agreement with the experimental data and the detailed CFD model was obtained, in terms of area-averaged deposition rates.

}
\normalfont\normalsize

\printnomenclature
\vskip .1em

\section{Introduction}
Fouling of solid surfaces and heat exchanger surfaces in particular, is a common and much studied problem in most process industries, as reflected in the review paper by \citet{MullerSteinhagen11}. 
Fouling is defined as the unwanted accumulation of solid (or semi-solid) material on solid surfaces. 
A similar phenomenon is the desired accumulation of solids e.g. in chemical vapor deposition \citep{Krishnan94,Kleijn89}. 
A common and costly problem in many industrial applications is the direct precipitation of super saturated fluids on heat exchanger surfaces. 
Typical examples are found in e.g. the high-temperature off-gas from waste incineration, metal production, or in power plants, where efficient heat recovery is key to sustainable production, and where a combination of direct precipitation and deposition of e.g. solid metal oxides is a major showstopper. 
Similar issues can be found in almost all process industries, and in the current work we study the deposition of a low-solubility salt (calcium carbonate, $CaCO_3$) from liquid water. 
By precipitation, we understand all types of phase transitions from a fluid to a relatively denser phase, e.g. gas $\to$ liquid (condensation), gas $\to$ solid (sublimation), liquid $\to$ solid (solidification). 
For some materials, the precipitate may have a crystalline structure (crystallization)(e.g. $CaCO_3$).

In our modelling work, fouling due to mass deposition from a fluid phase is grouped into two different classes; 1) particulate fouling, where particles carried by the fluid phase penetrate through the laminar boundary layer and stick to the wall (e.g. precipitates, dust, or soot particles) \citep{Johansen91,Johnsen09}; and 2) direct precipitation where the fluid is super-saturated close to the wall and a phase-transition occurs at the wall (current paper). 
The direct precipitation on solid surfaces is due to the molecular diffusion through the stagnant boundary layer close to the wall. 
This is a complex physical process where the diffusion flux of each species is coupled to the diffusion fluxes and thermodynamic/chemical properties of all the species present.
Commonly, a combination of 1 and 2 takes place. Fouling can only occur if the adhesive forces between the foulant and the wall are strong enough to overcome the flow-induced shear forces at the wall.

In previous papers, we developed frameworks for the mathematical modelling of particle deposition and re-entrainment \citep{Johansen91,Johnsen09} and direct precipitation \citep{Johnsen15}. 
In presentations \citep{Johnsen10,Johnsen16}, it was demonstrated how these models could be employed as wall boundary conditions (mass sinks) for CFD models. 
\citet{Paakkonen16} compared CFD simulations with experimental results with respect to $CaCO_3$ deposition in a lab-scale heat exchanger set-up. 
In the current paper we apply the wall function approach published in \citep{Johnsen15}, in a coarse grid CFD model, and test it against the detailed CFD modelling results and experimental data obtained by \citet{Paakkonen12,Paakkonen16}.

\section{Experimental Setup}
The modelling results are validated against experimental data from crystallization fouling on a heated surface. 
The experimental setup includes a flow-loop with a test-section (a rectangular flow channel), with ohmically heated test surfaces. 
In the present work, we investigate the case where the wall heat flux was a constant $q_w=52.5\nicefrac{kW}{m^2}$. 
A water-based test liquid, supersaturated with respect to $CaCO_3$, is circulated from a mixing tank and is filtered before entering the test section (average inlet velocities ranging from $u_{f,x,in}=0.2-0.4\nicefrac{m}{s}$ and temperature of $T_{in}=303K$), where $CaCO_3$ precipitates and deposits on the heated test surface. 
The growth of the fouling layer is monitored by measuring the temperature at the test surface. 
The decreased overall heat transfer coefficient due to the fouling layer (fouling resistance) will cause the test-section surface temperature to increase. 
Details of the experimental setup, procedure and results were described by \citet{Paakkonen12}.

\section{Model Description}
In the present paper, CFD is used to model experiments performed in the aforementioned experimental setup. 
Two different modelling approaches are employed; 1) \emph{Two-step fouling model} \citep{Paakkonen16}; and 2) \emph{Fouling wall function} for direct precipitation fouling \citep{Johnsen15,Johnsen16}. 
These two differ fundamentally in the way they approach the problem. 
Model 1 relies on a detailed CFD mesh close to the wall in order to be able to model the boundary layer phenomena correctly, and employs the traditional two-step approach (see e.g. \citep{Mullin01}) to model the deposition rate. 
Model 2, on the other hand, relies on a relatively coarse mesh, where the cell centers of the cells residing at the wall are in the log-layer. 
This approach employs a subgrid model to calculate the deposition rates from a set of simplified governing equations. 
For more details, see descriptions below as well as mentioned references. 
The main objective of the current paper is to shed light on the applicability of the wall function approach, since the successful application of such a method would be an essential step towards the cost-efficient modelling of many industry scale applications.

\InsFig{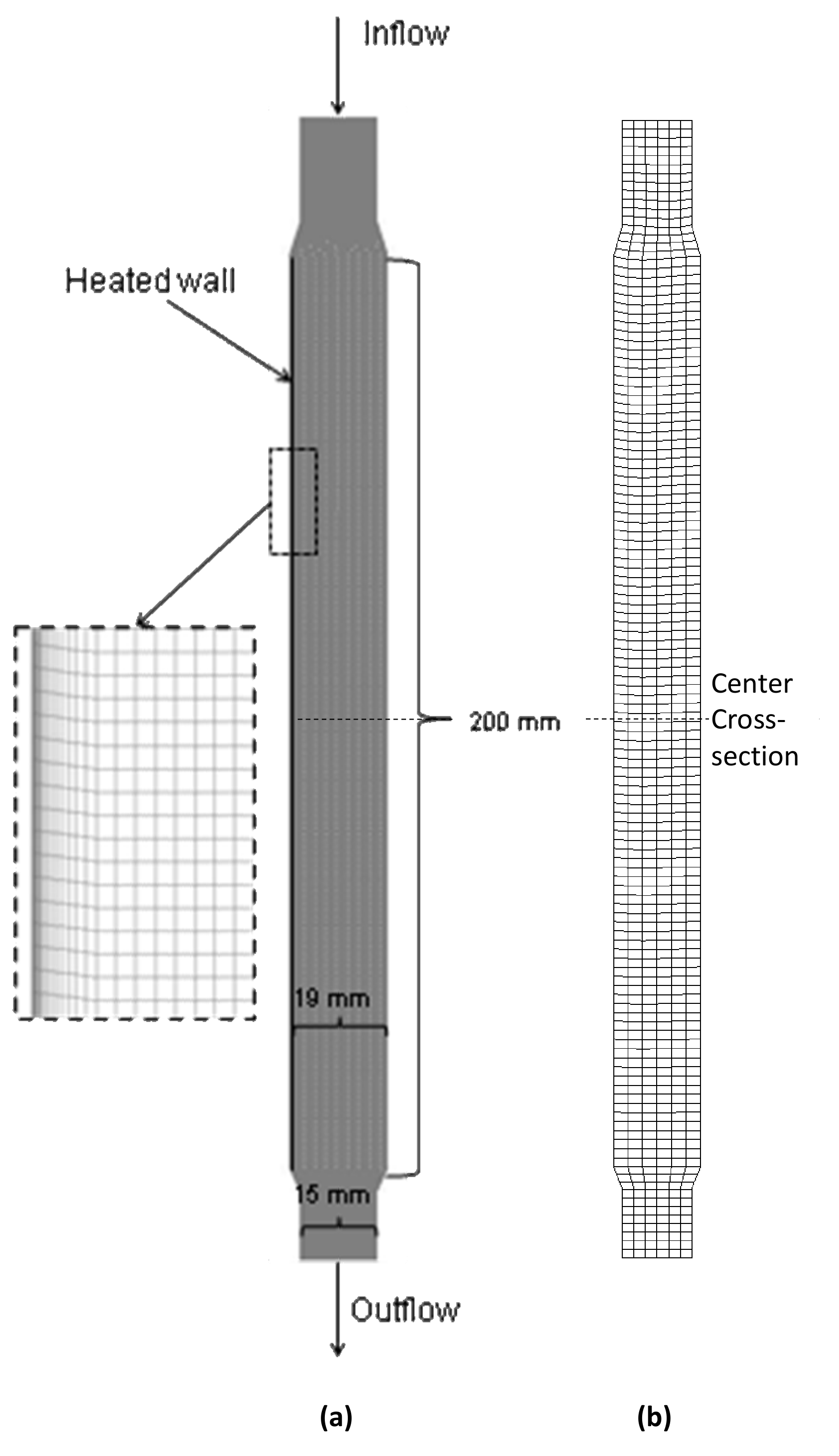}{Computational geometry and fine-mesh (a), for two-step fouling model \citep{Paakkonen16}, and coarse mesh (b), for fouling wall function model.}{Mesh}

\subsection{Geometry and Computational Mesh}
\Fig{Mesh} presents the 2D geometry used in the CFD simulations. 
The figure shows a 2D-representation of the liquid-filled gap between two parallel, vertical heat transfer surfaces. 
The liquid enters from the top and exits through the bottom. 
For more details about the experimental set-up, refer to \citet{Paakkonen12}.

Two different meshes were applied for the two different modelling approaches described below; namely a fine mesh, as shown in \Fig{Mesh}a \citep{Paakkonen16}, and a coarse mesh as shown in \Fig{Mesh}b. 
The coarse mesh was used with the wall function model \citep{Johnsen15} whereas the fine mesh was used with the two-step model \citep{Paakkonen16}.  
With the fine mesh, the $y^+$ value at the surface is about $0.08$, and the total number of cells is $76 000$. 
In the coarse mesh, the $y^+$ value at the wall is between $20$ (for the $u_{f,x,in}=0.2\nicefrac{m}{s}$ case) and $36$ (for the $u_{f,x,in}=0.4\nicefrac{m}{s}$ case), and the total number of cells is $276$. 
In addition, the wall function utilizes a 1-dimensional, logarithmic subgrid consisting of $300$ computational nodes, the first node at a wall-distance equal to $\nicefrac{1}{10 000}$th of the distance to the cell center in the coarse CFD mesh ($\sim2.34\cdot10^{-7}m$).

\subsection{Model Fluid}
The test liquid in the experiments was a mixture of various salts dissolved in water. 
Refer to \citep{Paakkonen15} for details. 
In the current modelling work it was assumed that the test fluid was a pure calcium carbonate, $CaCO_3$, solution in water. 
Thus, the mixture was considered as a dilute, electrically quasi-neutral ideal mixture with no chemical reactions. 
In the present paper, the $CaCO_3$ mass-fraction of $4.197\cdot {{10}^{-4}}{kg}/{kg}\;$ was used for the test fluid entering the model geometry. 
Temperature-dependent fluid properties (mass density, viscosity, diffusivity) were modelled in accordance with Table 2 in \citep{Paakkonen15}.

\subsection{Fouling Models}
Traditionally mass deposition at the wall surface, in crystallization fouling, is modelled based on a two-step approach. 
In the two-step modelling approach, the fouling process consists of 1) transport from the bulk to the vicinity of the wall, and 2) surface integration (i.e. adsorption onto the fouling layer).
The species transport to the vicinity of the crystal-fluid interface, is based on the difference between the bulk and interface concentrations. 
The mass transfer coefficient is typically estimated from empirical correlations. 
At the surface, the integration of the species into the crystal body is modelled as a pseudo chemical reaction driven by the difference between the interface and saturation concentrations. 
When the two steps are combined, the interfacial concentration, which is often unknown, cancels out of the model. 
The two-step approach has been used as a stand-alone model \citep{Bansal08,Helalizadeh05,Augustin95} as well as part of a CFD model \citep{Mwaba06a,Brahim03}.

\subsubsection{Two-step fouling model}
\citet{Paakkonen16} implemented the two-step model into CFD by utilizing the ability of CFD to model the transport of species to the vicinity of the surface, and thus provide the interfacial concentration difference between the surface and the fluid. 
To account for the wall shear-stress dependency of the adhesion probability seen in experiments \citep{Paakkonen15}, a time scaling factor was included in the model to scale the fluid residence time at the wall.

The mass deposition rate to the surface, based on the two-step approach, including the effect of the residence time \citep{Paakkonen15} can be expressed as
\begin{multline}\label{eq:twostep}
j_{dep}=\beta\left[\frac{1}{2}\left(\frac{\beta\rho_fu_\tau^2}{k'_r\mu_f}\right) + \left(C_b-C_{Sat}\right)-\right. \\ 
\left.-\sqrt{\frac{1}{4}\left(\frac{\beta\rho_f u_\tau^2}{k'_r\mu_f}\right)^2 + \frac{\beta\rho_f u_\tau^2}{k'_r\mu_f}\left(C_b-C_{Sat}\right)}\right]~.
\end{multline}
From the experiments, it was determined that the fouling process was controlled by surface integration \citep{Paakkonen12}. 
Thus, \Eq{twostep} reduces to
\begin{equation}
	j_{dep}=k'_r\left(C_b-C_{Sat}\right)^2\frac{\mu_f}{\rho_fu_\tau^2}~,
\end{equation}
where the rate constant for the surface integration can be determined from
\begin{equation}
	k'_r=k_0\exp{\left(\nicefrac{-E_a}{\mathcal{R}T}\right)}~.
\end{equation}
The pre-exponential factor $k_0=1.62\cdot10^{22}\nicefrac{m^4}{kgs^2}$, and the activation energy $E_a=148\nicefrac{kJ}{mol}$ were determined from the experiments, for the surface integration controlled fouling process \citep{Paakkonen15}.
The two-step fouling model was implemented into CFD as mass and momentum sink terms.  

\subsubsection{Fouling wall function}
The core idea of the fouling wall function approach is to formulate the species transport equations on one-dimensional form by applying appropriate approximations and simplifications in the turbulent boundary layer. 
Next, the simplified governing equations are solved on a local subgrid for each grid cell residing at the wall, to obtain the cell-specific deposition mass flux. 
Thus, the calculated species mass fluxes, at the wall, can be used as mass sinks in the CFD grid cells next to the wall. 

The set of steady-state governing equations consists of the Advection-Diffusion equation (ADE) for each species,
\begin{equation}
	\bVec{\nabla}\boldsymbol\cdot\left(\rho_fX_i\bVec{u}_f\right) + \bVec{\nabla}\boldsymbol\cdot\bVec{j}_{d,i} = 0~,
\end{equation}
the fluid mixture momentum and energy equations,
\begin{equation}
	\bVec{\nabla}\boldsymbol\cdot\left(\rho_f\bVec{u}_f\bVec{u}_f\right) = -\bVec{\nabla}P + \bVec{\nabla}\bVec{\tau} + \rho_f\bVec{g}~,
\end{equation}
\begin{equation}
	\bVec{\nabla}\boldsymbol\cdot\left(\rho_fh_{sens,f}\bVec{u}_f\right) = \bVec{\nabla}\left(k_f\bVec{\nabla}T\right) - \bVec{\nabla}\left(\sum_i{\bVec{j}_{i,d}h_{sens,i}}\right)~,
\end{equation}
and the restriction that the mass- and mole-fractions must sum to unity,
\begin{equation}
	\sum_{i}{X_i}=\sum_i{z_i}=1~.
\end{equation}

Introducing turbulence, dimensionless variables and appropriate simplifications, the simplified governing equations are obtained:
\begin{equation}
	\partial_{y^+}\left[\frac{\nu _t^+}{Sc_t}\rho_f^+\partial_{y^+}X_i\right]+\partial_{y^+}j_{d,i,y}^+=0
\end{equation}
gives the mass-fraction profiles; 
\begin{equation}
	\partial_{y^+}u_{f,x}^+=\nicefrac{1}/{\left(\mu^+ + \mu _t^+\right)}
\end{equation}
gives the dimensionless axial fluid mixture velocity profile; and 
\begin{equation}\label{eq:SimpEnEq}
	\partial_{y^+}\left[K_{(0)}^+T^+ + K_{(1)}^+\partial_{y^+}T^+ \right]=0
\end{equation}
gives the dimensionless temperature profile. 
\begin{equation}
	K_{(0)}^{+}\equiv \left( k_{f,c}^{+}+k_{f,t}^{+} \right)\left( {{\partial }_{{{y}^{+}}}}\ln c_{P}^{+} \right)-P{{r}_{w}}\sum\limits_{i=1}^{N}{_{i,d,y}^{+}c_{P,i}^{+}}~,
\end{equation}
and
\begin{equation}
	K_{(1)}^{+}\equiv k_{f}^{+}+k_{f,t}^{+}+k_{f,c}^{+}
\end{equation}
express the dimensionless groups in \Eq{SimpEnEq}. 
For more details, refer to \citep{Johnsen15}.

Due to the assumed weak effect of thermophoresis (due to small temperature gradients) and the lack of good estimates of the thermophoretic diffusivity, only diffusiophoresis (concentration gradient diffusion) was considered in the current work. 
Furthermore, it was assumed that the model fluid could be treated as a dilute, ideal mixture. 
This reduces the Maxwell-Stefan diffusion model to the Fickian diffusion model. 
The mixture mass density and viscosity was modelled in accordance with \citep{Paakkonen15}, while constant mixture thermal conductivity and specific heat capacity $0.6637\nicefrac{W}{m^2}$ of and $4182\nicefrac{J}{kgK}$, respectively, were used. 
The turbulent Schmidt number was set to $1$. 
The Maxwell-Stefan binary diffusivity was tuned so that the area averaged deposition rate matched that of the experiments, for the $u_{f,x,in}=0.2\nicefrac{m}{s}$ data-point, and was kept constant for the other inlet velocities. 
This resulted in a Fickian diffusivity of $3.64\cdot10^{-5}\nicefrac{m^2}{s}$.

\subsection{CFD Models}
CFD modelling was performed using the ANSYS FLUENT 16.2 CFD software.
Turbulence is modelled with the standard $k-\epsilon$ turbulence model. 
In the fine-mesh CFD model, the Enhanced Wall Treatment is employed to resolve the near wall region in the fine mesh model. 

Temperature dependent fluid properties were implemented via user-defined functions (UDFs) in accordance with \citep{Paakkonen15}. 
The fouling models were also implemented via UDFs and hooked into ANSYS Fluent via the adjust function hook. 
Due to the low deposition rates observed, it was expected that the mass transfer to the wall would have a very small effect on the bulk conditions in the coarse mesh. 
Thus, the fouling wall function was not utilized as a mass source, but was run on a frozen flow field.

     \begin{table}[!tb]
  \centering
  \caption{Wall ${{y}^{+}}$ values at the centre cross-section, for selected coarse grids with uniform node spacing, for inlet velocities $0.2$ and $0.4{m}/{s}\;$.}
  \label{tab:WallYp}
    \begin{tabular}{lcccc}
      \toprule
      \bf{No. of cells across channel} & $\bm4$ & $\bm6$ & $\bm8$ & $\bm{10}$ \\
	  \bf{Inlet velocity} & & & & \\
	  \midrule
	  $\bm{0.2}\nicefrac{\bm{m}}{\bm{s}}$ &	$30$ &	$20$ &	$15$ &	$11$ \\
	  $\bm{0.4}\nicefrac{\bm{m}}{\bm{s}}$ &	$49$ &	$36$ &	$25$ &	$20$ \\
    \bottomrule
    \end{tabular}
\end{table}

\InsFig{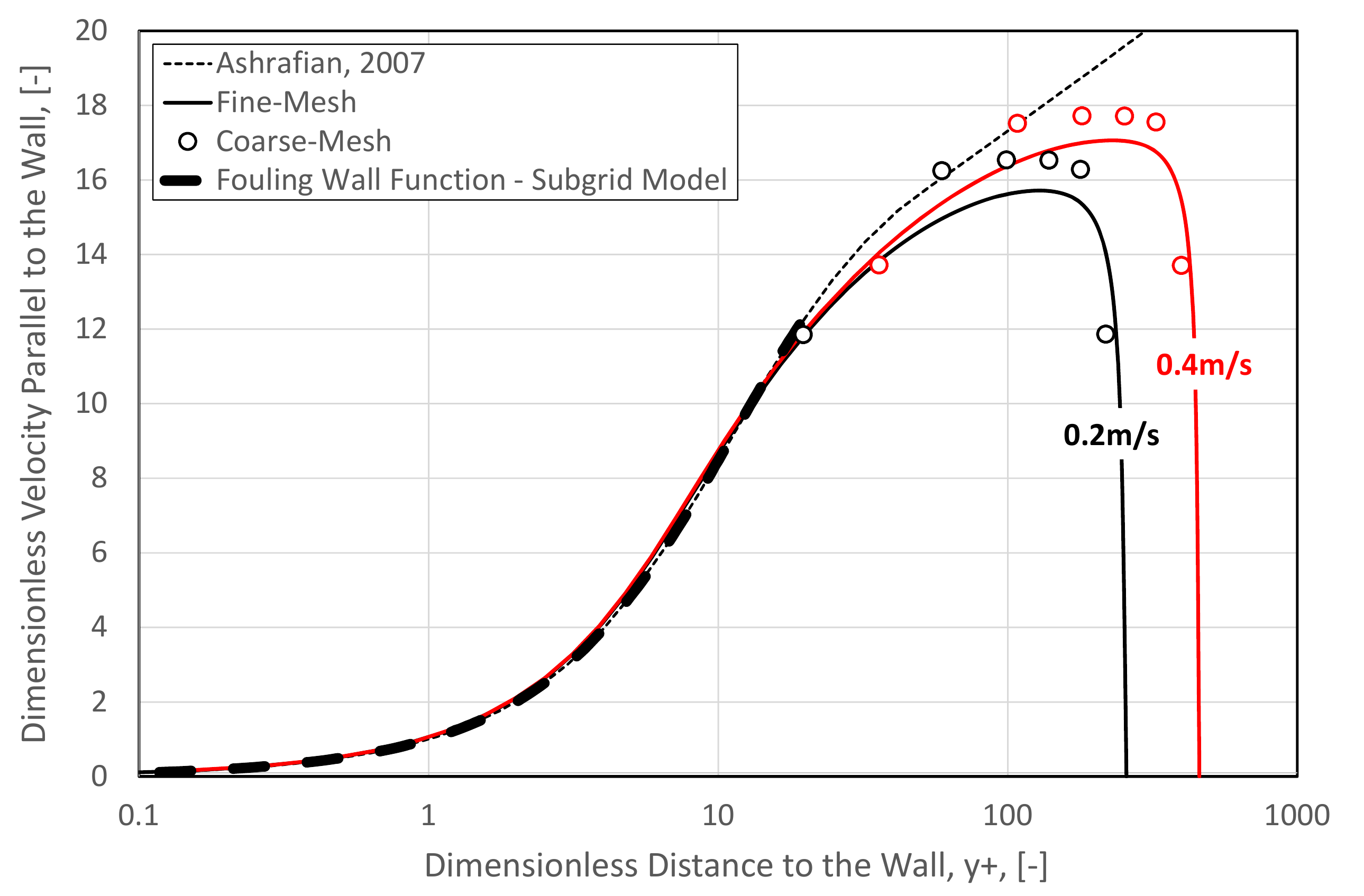}{Comparison of dimensionless velocities as functions of dimensionless wall distance at the center cross-section (isothermal conditions), for the coarse- (circles) and fine-mesh (solid lines) CFD models, the fouling wall function subgrid model (dashed, black line), and theoretical velocity profile \citep{Ashrafian07} (dotted, black line).}{DimLessVelProf}

\begin{figure*}[!bt]
  \centering
  \subfloat[]{\includegraphics[width=0.49\linewidth]{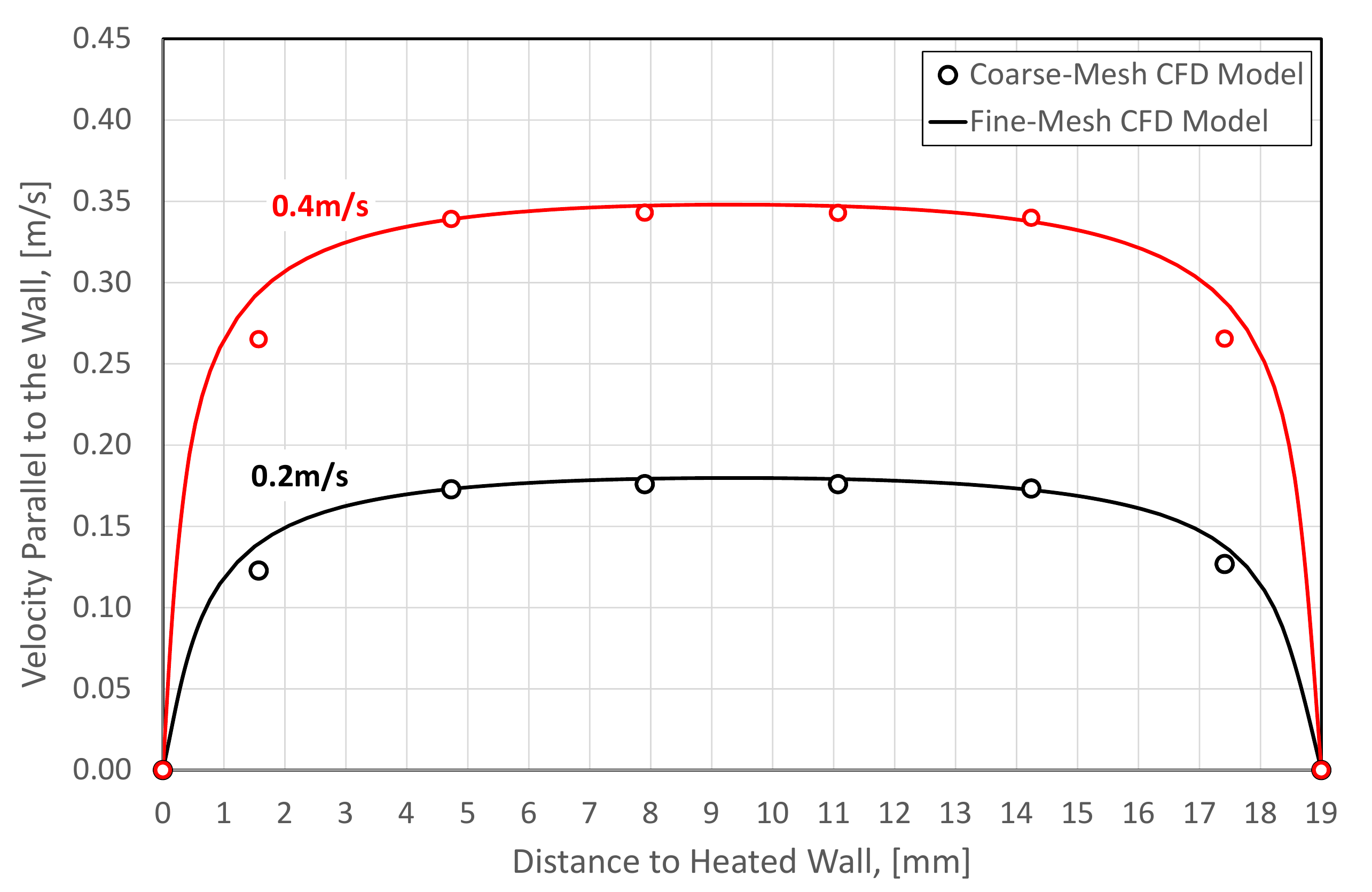}\label{fig:VelProf}}
  \subfloat[]{\includegraphics[width=0.49\linewidth]{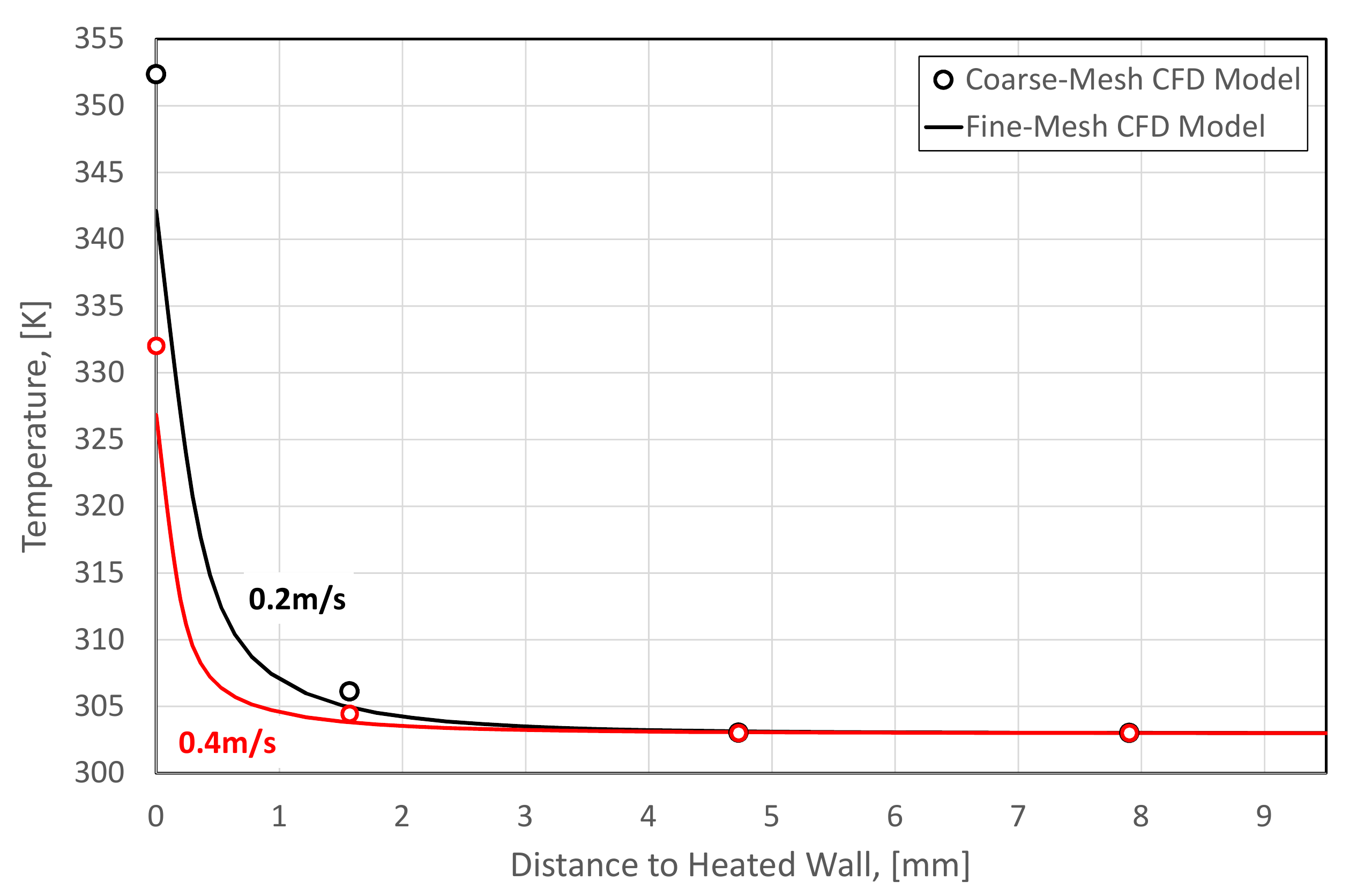}\label{fig:TempProf}}
 \caption{Comparison of parallel-to-wall flow velocity profiles (a) and temperature profiles (b) at the center duct cross-section ($x=100mm$), for coarse and fine-mesh CFD models, for wall heat flux of $52.5{kW}/{{{m}^{2}}}\;$ and inlet velocities $0.2{m}/{s}\;$ and $0.4{m}/{s}\;$.}
\label{fig:VelTempProf}
\end{figure*}

\subsection{Coarse-Mesh Velocity Wall Function}
The fouling wall function was designed to work on grids where the grid cells residing on the wall are in the log-layer. 
The main reason for this is that its bulk boundary conditions were chosen to be valid for fully developed turbulent flow. 
In the current experimental set-up, however, due to the low Reynolds numbers, such a stringent requirement of the wall $y^+$ left us with very coarse meshes. 
In \Tab{WallYp}, the approximate wall $y^+$ value at the center cross-section (see \Fig{Mesh}) is shown for various coarse meshes where the node spacing is constant across the channel. 

In order to predict the wall shear stress and general velocity profile accurately, on the coarse mesh, the wall function proposed by \citet{Ashrafian07}, was employed;
\begin{equation}
	u_{f,x}^{+}({{y}^{+}})=\left\{ \begin{matrix*}[l]
   11.4\arctan \left( \frac{{{y}^{+}}}{11.4} \right)\text{,} & {{y}^{+}}\le y_{*}^{+}  \\
   \frac{1}{\kappa }\ln \left( \frac{1+\kappa {{y}^{+}}}{1+\kappa y_{*}^{+}} \right)+u_{f,x}^{+}(y_{*}^{+}), & {{y}^{+}}>y_{*}^{+}  \\
\end{matrix*} \right.~,
\end{equation}
with the dimensionless turbulent kinematic viscosity
\begin{equation}
	\nu _{t}^{+}=\left\{ \begin{matrix*}[l]
   {{\left( \frac{{{y}^{+}}}{11.4} \right)}^{2}} & {{y}^{+}}\le y_{*}^{+}  \\
   \kappa {{y}^{+}} & {{y}^{+}}>y_{*}^{+}  \\
\end{matrix*} \right.~,
\end{equation}
where dimensionless velocity is defined as $u_{f,x}^{+}={u_{f,x}^{{}}}/{{{u}_{\tau }}}\;$, dimensionless wall distance is defined as ${{y}^{+}}={{{u}_{\tau }}y}/{\nu }\;$, $y_{*}^{+}=51.98$, and $\kappa =\text{0}\text{.42}$ is the von K\'arm\'an constant.

A sensitivity study was done to investigate how the coarse meshes performed against the fine-mesh CFD model and the Ashrafian-Johansen wall function. 
It was determined that the mesh with $6$ cells across the channel reproduced the fine-mesh velocity and temperature profiles quite well and at the same time gave an acceptable wall ${{y}^{+}}$ value. 
In \Fig{DimLessVelProf}, it is shown how the fine-mesh and coarse-mesh CFD models perform against the profile published by \citet{Ashrafian07} under isothermal conditions (no heating), in terms of dimensionless variables. 
The deviations at high ${{y}^{+}}$ values are due to the effect of the opposing channel wall and the relatively low Reynolds numbers investigated. 
For the coarse-mesh CFD model, the fouling wall function subgrid model is included for validation of the subgrid velocity profile. 
In \Fig{VelProf}, the coarse and fine-mesh axial velocity profiles at the center cross-section are compared, and in \Fig{TempProf}, the temperature profiles are compared. It can be seen that generally, the axial velocity was underpredicted, in the coarse-mesh CFD model, whereas the temperature was overpredicted.

\subsection{Boundary Conditions for the Fouling Wall Function}
The fouling wall function requires boundary conditions for temperature and species mass-fractions at the wall as well as axial velocity, temperature and species mass-fractions in the bulk. 
The bulk values as well as the wall temperature are taken directly from the CFD model via the inbuilt macro library in ANSYS Fluent, and utilized as Dirichlet boundary conditions in the subgrid model. 
The species-specific mass-fraction boundary conditions at the wall, however, require special attention. 
First, the type of boundary condition depends on whether the species is depositing or not; second, they depend on which diffusive transport mechanisms are dominating close to the wall \citep{Johnsen17}.

E.g., consider the case where diffusion due to mass-fraction gradients (diffusiophoresis) is the sole transport mechanism close to the wall. For the non-depositing species, the mass-fraction gradient at the wall must be zero to ensure zero deposition flux, and we employ the Neuman BC for the ADE, at the wall. 
For the depositing species, however, we do not have a priori knowledge of the deposition flux, so we cannot use the mass-fraction gradient as a BC. 
We have to use the Dirichlet BC. 
That is, we need to specify the mass-fractions of the depositing species, at the wall.

The mass-fractions at the wall (interface mass-fractions) are consequences of the balance between transport through the turbulent boundary layer and the species integration into the crystal lattice. 
Therefore, it is a function of e.g. temperature, temperature gradient, composition, composition gradients, wall shear stress, crystal properties, etc. 
Thus, the interface mass-fraction is not just a fixed boundary condition, but is in fact part of the solution itself. 
If the kinetics of the surface reaction are known, it is possible to estimate the interface mass-fractions. 
Then, an iterative procedure can be employed to find the interface mass-fraction that ensures that the transport rate through the boundary layer and the integration rate into the crystal are identical \citep{Johnsen17}. 

Lacking accurate predictions of the surface reaction rates, the current wall function model employed interface concentrations obtained from the fine-mesh CFD model (see \Fig{Xivsx}). 
These concentrations are dependent on both wall temperature and inlet velocity (wall shear stress). 
By curve fitting the Logistic function,
\begin{equation}\label{eq:Xireg}
	{{X}_{I,reg}}=\frac{a}{1+{{\left( {{{T}_{w}}}/{b}\; \right)}^{c}}}\text{  }~,
\end{equation}
to the fine-mesh CFD data, we obtained good representations of the interface mass-fractions for each inlet velocity case. 
The inlet velocity-dependent fitting parameters, $a$, $b$, and $c$, are shown in \Fig{FitParam}, and could be accurately described in terms of 3rd and 2nd order polynomials; 
\begin{eqnarray}
	a&=&{{a}_{0}}+{{a}_{1}}{{u}_{f,x,in}}+{{a}_{2}}u_{f,x,in}^{2}+{{a}_{3}}u_{f,x,in}^{3}~,\label{eq:regparama}\\
	b&=&{{b}_{0}}+{{b}_{1}}{{u}_{f,x,in}}+{{b}_{2}}u_{f,x,in}^{2}~,\label{eq:regparamb}\\
	c&=&{{c}_{0}}+{{c}_{1}}{{u}_{f,x,in}}+{{c}_{2}}u_{f,x,in}^{2}~.\label{eq:regparamc}
\end{eqnarray}
The coefficients are given in \Tab{FitParam}. 
Employing \Eq{Xireg} with coefficients given by Eqs. \ref{eq:regparama}-\ref{eq:regparamc}, we got a good, general representation of the CFD-data (see black circles in \Fig{FitParam}).

\begin{table}[!tb]
  \centering
  \caption{Curve fit polynomial coefficients for velocity dependence of interface mass-fractions (Eqs. \ref{eq:regparama}-\ref{eq:regparamc}).}
  \label{tab:FitParam}
    \begin{tabular}{
  l *{3}{S[table-format=-1.3,table-space-text-post=***]}
}
    \toprule
	 		&	{$\bm a$}	&	{$\bm b$}	&	{$\bm c$}\\
	 \midrule
	$\bm 0$	&	0.251654	&	342.436		&	409.600	\\
	$\bm 1$	&	 1.28476	&	26.6133		&	-2179.69\\
	$\bm 2$	&	-3.57731	&	-112.872	&	4968.05	\\
	$\bm 3$	&	 3.41471	&				&			\\
    \bottomrule
    \end{tabular}
\end{table}

\InsFig{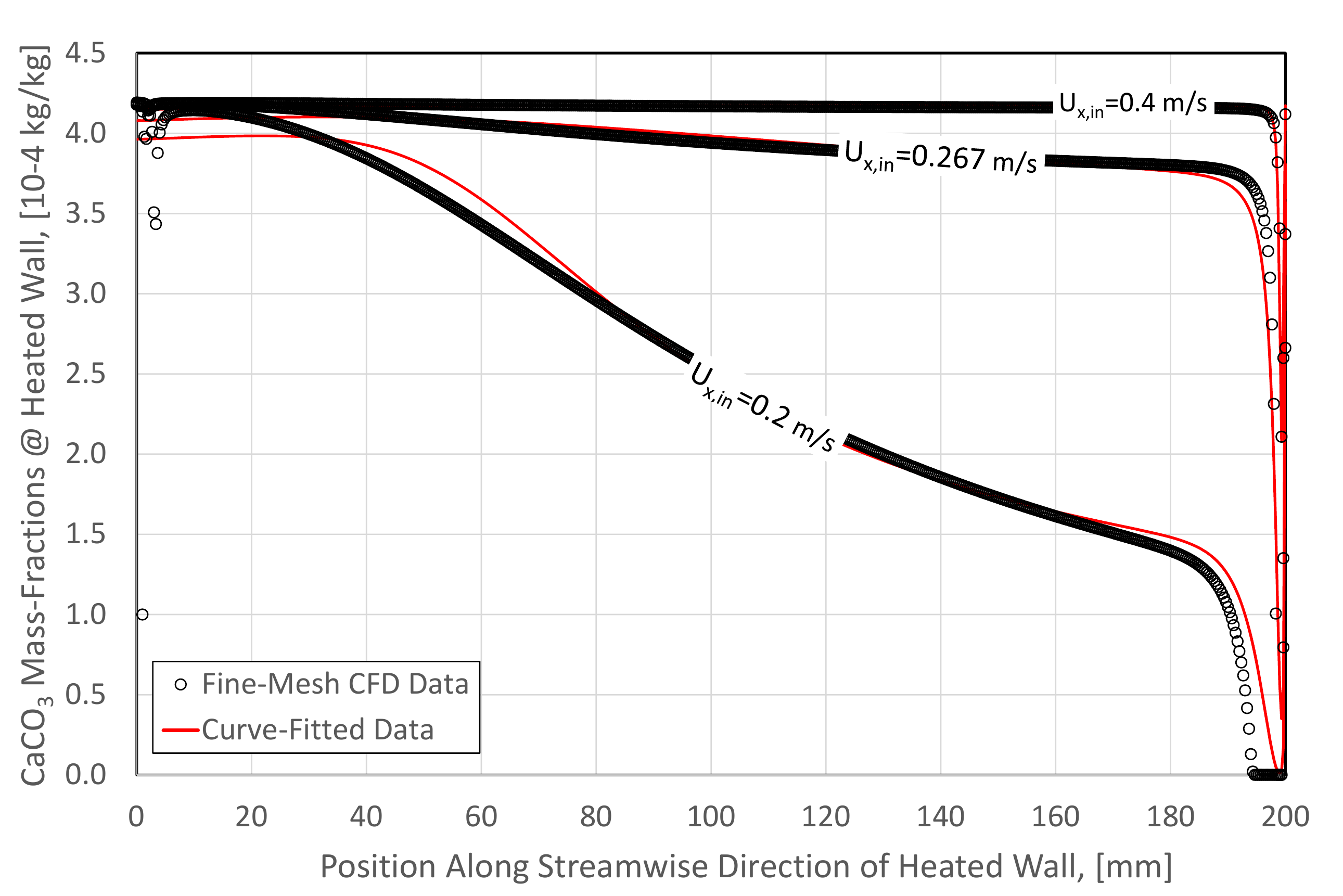}{Comparison of fine-mesh CFD (black circles) and best-fit (red lines) $CaCO_3$ mass-fractions at the wall plotted against the axial position along the heated wall, for selected inlet velocities.}{Xivsx}

\InsFig{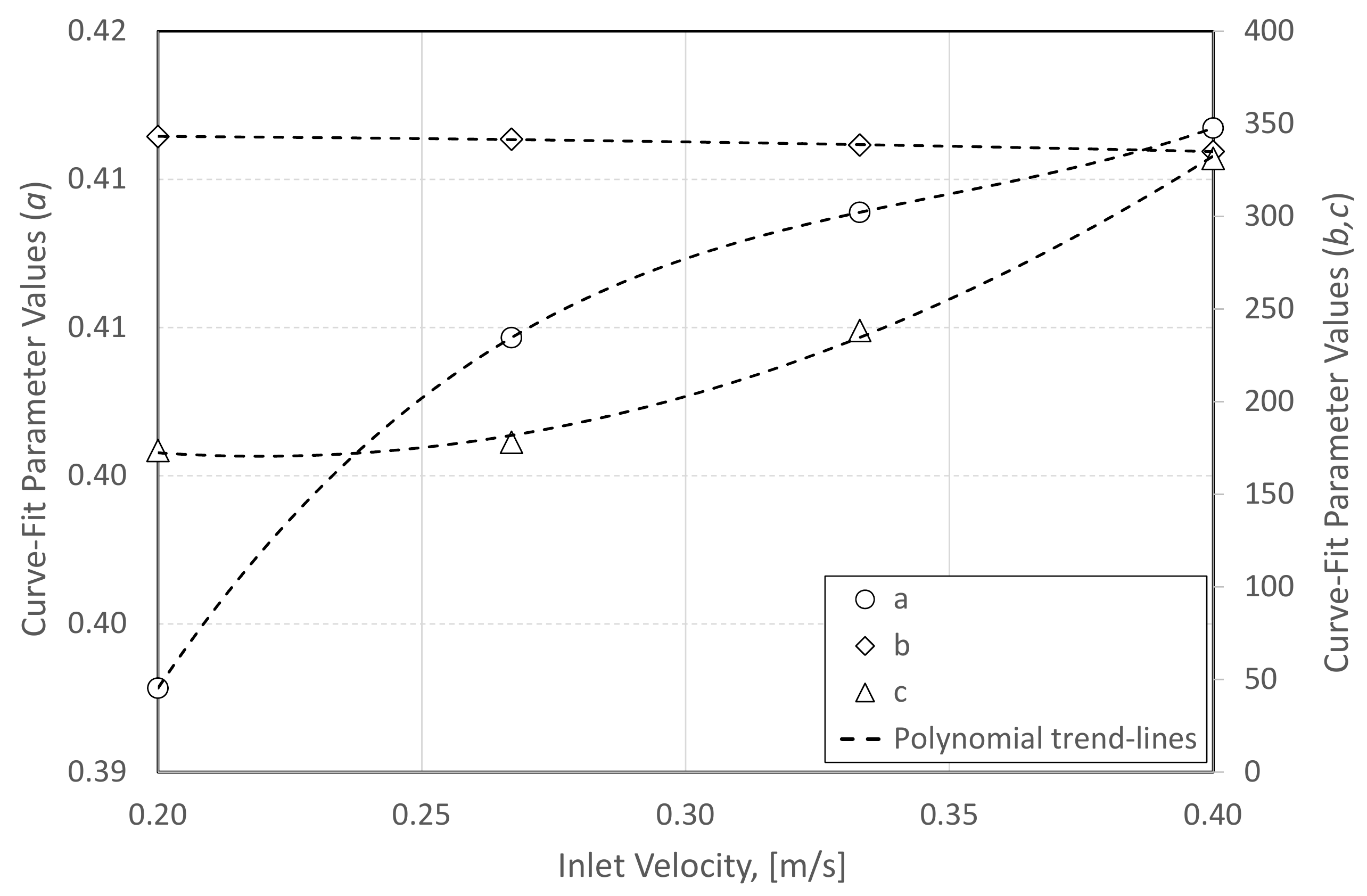}{Curve-fit parameters $a$, $b$, and $c$ (see \Eq{Xireg}) plotted as functions of inlet velocity, along with best-fit polynomial trend-lines (see Eqs. \ref{eq:regparama}-\ref{eq:regparamc} and \Tab{FitParam}).}{FitParam}

\InsFig{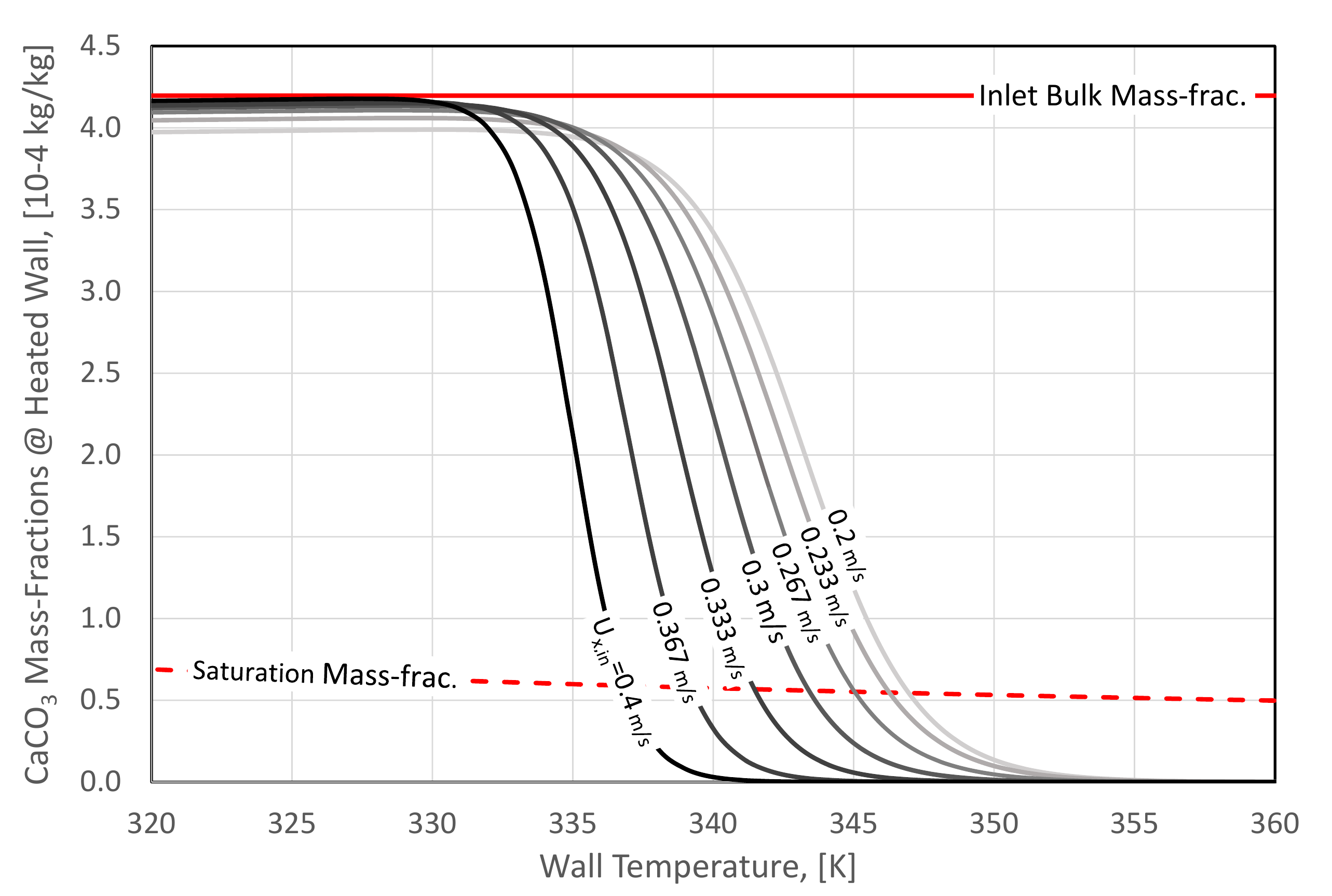}{Comparison of temperature dependency of $CaCO_3$ bulk (red line), saturation (dashed red line), and interface (calculated for various inlet velocities, from \Eq{Xireg}: black$=0.4\nicefrac{m}{s}$, light gray$=0.2\nicefrac{m}{s}$) mass-fractions, at the wall.}{XivsTw}

\InsFig{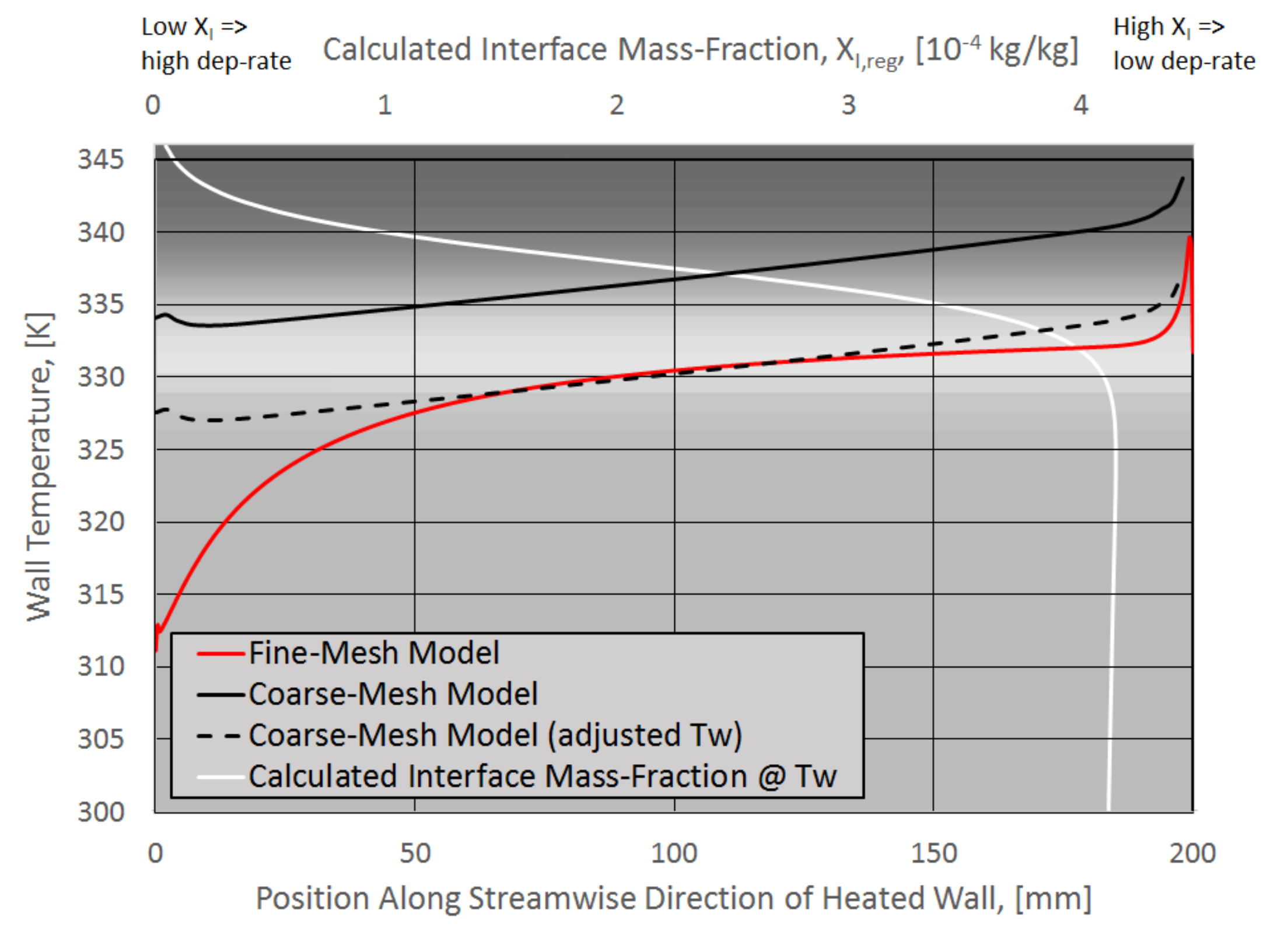}{Wall temperature vs. position along the heated wall, for the ${{u}_{f,x,in}}=0.333{m}/{s}\;$ case, for the fine-mesh model (red), coarse-mesh model (solid black), and adjusted coarse-mesh wall-temperature ($-6.5K$) (dashed black). The relationship between the wall temperature and the interface mass-fraction is shown by the white curve. The contour plot in the background corresponds to the interface mass-fraction values at given wall temperatures (dark gray corresponds to low ${{X}_{I}}$,  and light gray corresponds to high ${{X}_{I}}$) and links the modelled, local wall temperatures with an expected local interface mass-fraction.}{Twvsx}

\Fig{XivsTw} shows the temperature dependence of the calculated interface mass-fraction (\Eq{Xireg}), for selected inlet velocity cases. 
It is seen that the interface mass-fraction drops from close to the bulk value to zero, at a certain threshold temperature, which appears to be dependent on the inlet velocity. 
In reality, this is a consequence of the complex interplay between mass deposition rate, interface mass-fraction, wall temperature, and wall shear stress. 
We will be content, however, to consider this as an inlet velocity dependent feature. 
At temperatures below the threshold, the deposition regime is interface controlled, whereas at higher temperatures it is diffusion controlled. 
\citet{Paakkonen12} concluded that the fouling regime was interface controlled, in these experiments, since the over-all deposition rate is not increasing for increasing flow-velocities, as would be expected for a mass transfer controlled fouling regime. 
However, various segments of the heated wall may be in different fouling regimes depending on the local flow conditions and wall temperature, as indicated in \Fig{Twvsx}.
In general, the higher the difference between the bulk and interface mass-fractions, the higher the deposition rate (mind that at interface mass-fractions below the metastable equilibrium mass-fraction, fouling might not take place at all). 
Thus, the deposition rate at locations with wall temperatures above the threshold can be expected to dominate. 
Since the coarse-mesh CFD model is prone to overpredict the wall temperature, as was discussed above (see \Fig{TempProf}), there is a risk that the interface mass-fraction is severely underpredicted if the true wall temperature is lower than, but close to the threshold temperature. 
To reduce the risk of overprediction of deposition rates, a fixed $6.5K$ was subtracted from the wall temperature when calculating the interface mass-fraction from \Eq{Xireg}. 
\Fig{Twvsx} shows that a greater part of the overpredicted wall temperature curve (solid black) is in the low interface mass-fraction region (dark gray area) than the fine-mesh model wall temperature curve (red). 
Hence, a greater part of the wall will have low interface mass-fraction in the coarse-mesh model than in the fine-mesh model.
The corrected wall temperature curve (dashed black), however, is more similar to the fine-mesh model temperature curve.
Furthermore, interface mass-fractions below the saturation mass-fraction indicate that the fluid is undersaturated at the crystal surface. 
Physically this means that deposition is unfavorable with respect to minimizing the Gibbs free energy, thus no deposition will take place \citep{Johnsen17}. 
Therefore, the Dirichlet boundary condition for the $CaCO_3$ was set to
\begin{equation}
	{{X}_{I,CaC{{O}_{3}}}}=\max \left( {{X}_{I,reg}},{{X}_{Sat}} \right)~.
\end{equation}

\section{Results and Discussion}
The ambition in the current work was to demonstrate the applicability of a previously developed fouling wall function framework \citep{Johnsen15}, in practice. 
To approach this objective, its implementation, as a user-defined function in ANSYS Fluent 16.2, was employed to demonstrate how it performs against a more traditional two-step fouling modelling approach \citep{Paakkonen16}, in the context of a well-controlled laboratory experiment \citep{Paakkonen12}.

The main motivation for developing the fouling wall function was to eliminate the need to resolve the turbulent boundary layer and enable efficient fouling modelling in industry scale CFD simulations. 
Hence, the modelling framework relies on relatively high wall ${{y}^{+}}$values in the CFD cells residing at the wall. 
This proved to be challenging in the employment of the above cited experimental and simulation data, for comparison. 
Due to the low Reynolds numbers encountered in the data from P\"a\"akk\"onen et al., it was necessary to confide in a very coarse CFD mesh as basis for the fouling wall function modelling (see \Fig{Mesh}).

\InsFig{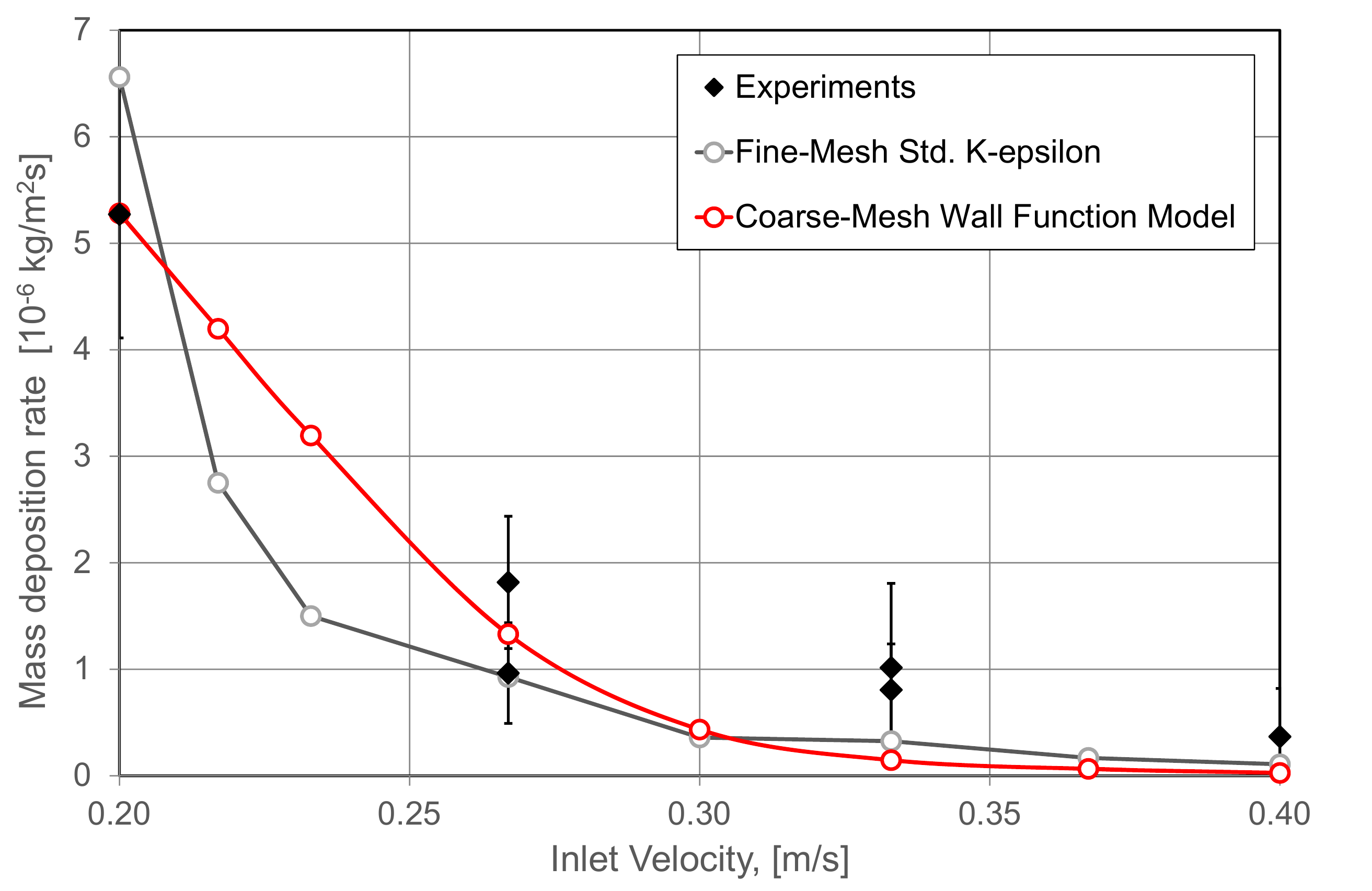}{Comparison of the area averaged mass deposition rates from the fine-mesh two-step model and the coarse-mesh fouling wall function model with the experimental data.}{AAvgDepRate}

The mass deposition rates predicted by the fouling wall function are depending directly on the wall function boundary conditions;
\begin{compactitem}
	\item wall mass-fractions for the depositing species,
	\item bulk and wall temperatures,
	\item bulk velocity parallel to the wall, 
\end{compactitem}
where \emph{bulk} refers to the center of the CFD grid cells residing at the wall. 
Thus, accurate prediction of the deposition rates rely heavily on the accurate CFD modelling of these quantities.

By utilizing the wall function published by \citet{Ashrafian07}, we managed to reproduce the fine-mesh CFD model velocity and temperature profiles fairly well, qualitatively. 
However, the quantitative discrepancy turned out to be the major source of error in the fouling wall function modelling results. 
In \Fig{DimLessVelProf}, it can be seen how the dimensionless velocity profiles are comparable in the absence of heating, and in \Fig{VelTempProf} it can be seen how dimensional velocity and temperature profiles are comparable under constant heating of $52.5{kW}/{{{m}^{2}}}\;$. 
The effect on the velocity profiles, by turning on/off heating was minimal.

Since thermophoresis was neglected in the current work, the role of the temperature was to provide the temperature dependent fluid properties (mass density, viscosity, and saturation mass-fraction), and interface mass-fraction for the depositing species. 
Although the inaccurate prediction of any of these will affect the predicted mass deposition rate to some extent, it seemed that the effect of the inaccurate prediction of the interface mass-fraction was the most severe. 
As was indicated in figures \ref{fig:XivsTw} and \ref{fig:Twvsx}, even modest errors in the local wall temperature could result in a severely miss-represented interface mass-fraction. 
Since the mass deposition rate is expected to scale approximately linearly with the difference between the bulk and interface mass-fractions, the mass deposition rate can be off by an order of magnitude by just a slight overprediction of the wall temperature, as seen in \Fig{XivsTw}. 
To avoid underpredicting the interface mass-fraction due to overprediction of the wall temperature, the temperature was subtracted a fixed $6.5K$ when calculating the interface mass-fractions (see \Fig{Twvsx}).


\Fig{AAvgDepRate} presents a comparison between the experimental data, the fine-mesh two-step fouling model data, and the data obtained from the coarse-mesh fouling wall function model. 
In the absence of reliable measurements/calculations of the diffusivity, it was treated as a calibration parameter, for the fouling wall function.
The fouling wall function data were thus obtained with a tuned diffusivity of $3.64\cdot10^{-5}\nicefrac{m^2}{s}$, reproducing the ${{u}_{f,x,in}}=0.2\nicefrac{m}{s}$ experimental data point.
The same, constant diffusivity was used in all grid cells along the wall, for all the inlet velocity cases. 
Despite the issues with predicting the required boundary conditions for the fouling wall function model accurately, the modelling results compared very well with the results from the fine-mesh two-step fouling modelling and the experimental data, in terms of the area averaged mass deposition rate. 

In \Fig{LocalDepRate}, the local deposition rates are compared for the fine-mesh two-step model and the coarse-mesh fouling wall function model. 
It can be seen that even if the area-averaged values compared well, the local values differs significantly. 
The mismatch seems primarily to be due to
\begin{compactitem}
	\item inaccurate prediction of interface mass-fraction;
	\item inaccurate prediction of wall temperature in the coarse mesh;
	\item inaccurate prediction of bulk velocity in the coarse mesh.
\end{compactitem}
The most crucial improvement to the fouling wall function model would be to get accurate interface mass-fractions. 
An in-depth study of these effects are left to future investigations. 
In the meantime, we are content to summarize that the fouling wall function approach performed very well in a scenario, slightly outside the design specifications of the modelling framework, with respect to the Reynolds number.

\InsFig{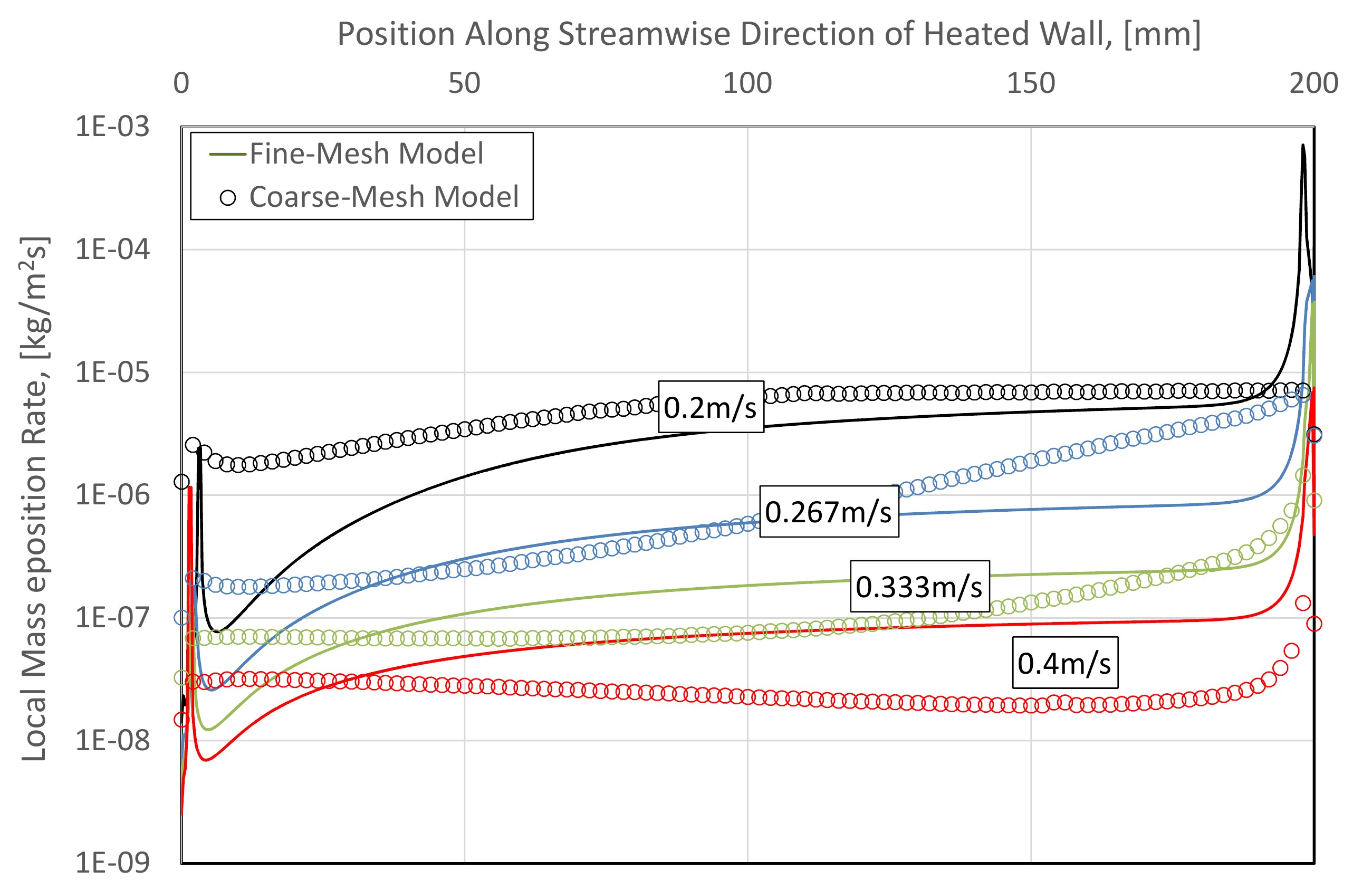}{Comparison of the local mass deposition rates calculated by the fine-mesh two-step model (lines) and the coarse-mesh fouling wall function model (circles), at selected inlet velocities.}{LocalDepRate}

The model fluid used in the current paper is a coarse simplification of the actual fluid employed in the cited experiments. 
The real fluid was a salt-water solution involving a multitude of chemically reacting ions and molecules. 
This is reflected by the fact that the content of dissolved $CaC{{O}_{3}}$ in the model fluid, is much higher than the saturation concentration. 
Thus, the modelled $CaC{{O}_{3}}$ may be seen as a pseudo-component representing e.g. the true $CaC{{O}_{3}}$ fraction in addition to $C{{a}^{2+}}$, $CO_{3}^{2-}$ and possibly other species. 
In the present case, at relatively low concentrations, this simplification seems to be justified in both modelling approaches employed. 
However, this may be part of the explanation of the local difference between deposition rates resulting from the two modelling methods.

The present demonstration case indicates that in industry-scale applications, where very fine meshes are infeasible, the wall function approach may provide a means to do physically detailed simulations of complex fluids, in complex geometries, at reasonable computational cost. 
In particular, if it can be assumed that the deposition rates are so small that they do not affect the flow field significantly, the savings in computational cost will be great. 
Then, the fouling wall function can be run on a frozen flow-field, and sensitivity studies or optimization studies on e.g. diffusivities, wall surface properties, etc., that does not affect the macro scale flow-fields can be performed without the need to update the frozen flow-field. 
Establishing the frozen flow-field on the coarse mesh, without the fouling wall function activated is very efficient due to the low number of computational cells needed. 
Then, running multiple fouling scenarios can be done on that flow-field just by changing input parameters to the fouling wall function and running one single CFD iteration, for each fouling scenario, with the fouling wall function activated.

\section{Conclusion}
Two different CFD modelling approaches were compared with experimental data on mass deposition rates in an experimental heat exchanger set-up. 
The two CFD strategies resolved the fine length-scales determining the mass transfer through the turbulent boundary layer, in two different ways: 1) the refinement was done in the 2D CFD mesh, resulting in a relatively high number of grid cells and a wall $y^+$ of ca. $0.08$; and 2) the refinement was taken into account in a wall function utilizing a 1-dimensional subgrid, allowing for a coarse CFD mesh with wall $y^+$ of about $30$. 
The fine-mesh CFD model utilized a traditional two-step modelling approach for the mass deposition modelling, complemented with the fluid residence-time at the wall, whereas the coarse-mesh CFD model wall function solved the coupled Advection-Diffusion, momentum and energy equations on a local subgrid to estimate the mass deposition rates. 

The coarse-mesh model performed very well compared to the fine-mesh model and experimental data, with respect to area average deposition rates. 
Significant mismatch was observed, however, in the local deposition rates. 
The lacking accuracy in the coarse-mesh model was mainly due to the challenges in predicting interface mass-fractions, wall temperatures and bulk velocities, on the very coarse mesh. 
	
The over-all good performance of the coarse-mesh model gives strong support to the idea that the wall function approach may provide a means to do physically detailed simulations of complex fluids, in complex, industry-scale geometries, at a reasonable computational cost.

\section{Acknowledgements}
This work was funded by the Research Council of Norway and The Norwegian Ferroalloy Producers Research Association, through the SCORE project \citep{SCORE}. 
Sverre expresses his gratitude towards the University of Oulu, Environmental and Chemical Engineering, FINLAND, for hosting him and his family during August 2016.

\balance
\bibliographystyle{CFD2017}
\bibliography{References}

\end{document}